\documentclass[journal]{IEEEtran}
\usepackage{graphicx}
\usepackage{epsf}
\usepackage{mathtools}
\usepackage{amssymb}
\usepackage{amsmath, amsfonts}
\usepackage{latexsym}
\usepackage{subfigure}
\usepackage{multirow}
\usepackage{multicol}
\usepackage[table]{xcolor}
\usepackage{color}

\usepackage{tabularx}
\usepackage{lipsum}

\usepackage{algorithm}
\usepackage{algorithmic}

\usepackage{mathrsfs}

\usepackage{fixltx2e}

\usepackage[mathscr]{euscript}

\usepackage{commath}
\usepackage{MnSymbol}

\usepackage{mathtools} 
\DeclarePairedDelimiterX\setc[2]{[}{]}{\,#1 \;\delimsize\vert\; #2\,}
\DeclarePairedDelimiterX\parth[2]{(}{)}{\,#1 \;\delimsize\vert\; #2\,}

\PassOptionsToPackage{hyphens}{url}
\usepackage[unicode=true, bookmarks=true, bookmarksnumbered=true, bookmarksopen=true, bookmarksopenlevel=1, breaklinks=false, pdfborder={0 0 0}, pdfborderstyle={}, backref=false, colorlinks=false]{hyperref}

\definecolor{orange}{RGB}{255,127,0}
\definecolor{blue}{RGB}{0,0,255}
\definecolor{red}{RGB}{255,0,0}
\definecolor{green}{RGB}{50,160,50}
\definecolor{grey}{RGB}{125,120,125}

\begin{document}
{
\title{{\fontsize{26}{10}\selectfont 5G or Wi-Fi for HA/DR in the 60 GHz Band?}}

\author
{
Md Fahad Kabir and Seungmo Kim, \textit{Member}, \textit{IEEE}

\thanks{M. F. Kabir and S. Kim are with the Department of Electrical and Computer Engineering, Georgia Southern University in Statesboro, GA, USA. The corresponding author is S. Kim: contact at seungmokim@georgiasouthern.edu.}
}

\maketitle
\begin{abstract}
This paper studies the feasibility of Wi-Fi and 5G technologies in the unlicensed 60 GHz band for humanitarian assistance and disaster relief (HA/DR) operations. Building a wireless communications system in the 60 GHz band can benefit HA/DR activities for two reasons: (i) no license is needed; (ii) a wide bandwidth is available. Our simulation results show that both of Wi-Fi and 5G can achieve data rates exceeding the requirements for most of the HA/DR missions, which proves the feasibility of the two wireless technologies for operation of high-data-rate HA/DR activities such as real-time video streaming.
\end{abstract}


\vspace{0.2 in}
\section{Introduction}\label{sec_intro}
For public safety, it is critical to have an option to provide and keep up communications amid and after a calamity or crisis \cite{fema}. In the case of an emergency or natural catastrophe, the telecommunications structure are likely to be damaged. Thus, wireless communications for humanitarian assistance and disaster relief (HA/DR) activities may become non-available, which may be significant threats to the society's safety. Further, for modern disaster management, higher data rates are required: examples include from video streaming for live footage and on-time response to disaster shelters experiencing instantaneous burst in bandwidth demand while accommodating large numbers of victims.

The 60 GHz band (57-71 GHz) has been attracting a large public interest ever since it was released by the Federal Communications Commission (FCC) in 2016 \cite{fcc}. The key benefits that the emergency communications can take from using the 60 GHz spectrum are two-fold: (i) it is an \textit{unlicensed} band in which any wireless system is allowed to operate without a license granted by the FCC; (ii) its historic \textit{abundance in the bandwidth} of 14 GHz enables a myriad of high-data-rate applications for emergency communications.

The key challenge in establishing emergency communications systems in the 60 GHz band is \textit{interference among dissimilar wireless systems}. Notice that any wireless system can operate without license in this unlicensed band. In fact, the Wireless Fidelity (Wi-Fi) and the 5th Generation Wireless (5G), the two most proliferating wireless technologies, already rolled out earlier versions of unlicensed systems operating at 60 GHz.

In this context, this paper investigates the \textit{feasibilities of 5G and Wi-Fi} in the 60 GHz band for HA/DR applications. Specifically, it studies the data rates achieved by 5G and Wi-Fi under coexistence with each other. From extension of a recent work \cite{lu2017downlink}, we focus on finding the exact number of access points (APs) at which a sufficient data rate is achieved for HA/DR activities.

\begin{figure*}[t]
\centerline{\includegraphics[width = \textwidth]{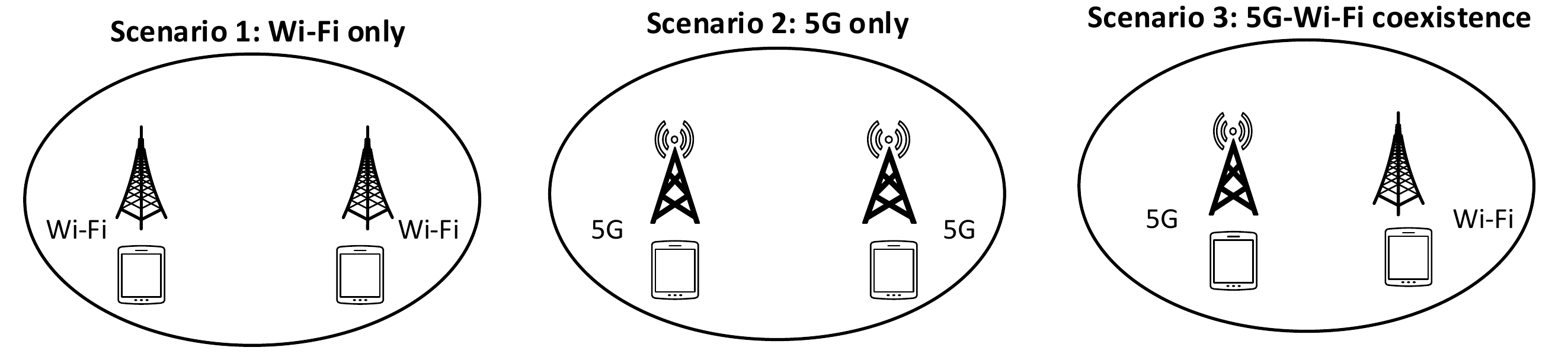}}
\caption{Operation scenarios of 5G and Wi-Fi in the 60 GHz}
\label{fig_pairs}
\vspace{0.1 in}
\end{figure*}

\section{Related Work}\label{sec_related}
Initial work had discussed the significance of the 60 GHz band and key research challenges \cite{yang06}\cite{rappaport}. Ever since the earliest discussions, identification of technologies and issues for deploying next-generation short-range wireless networks still remains as an open problem mainly due to the inter-technology interference issues. So far, the consensus has been focusing on three technologies: Wi-Fi, cellular (\textit{e.g.}, 5G), and licensed assisted access (LAA) such as Long-Term Evolution Unlicensed (LTE-U).

As such, one needs to precisely characterize the coexistence and interoperability among the dissimilar wireless technologies. Nevertheless, only few prior studies focused on addressing the coexistence issue. Therefore, this paper identifies related research on cross layer has also been conducted in order to further optimize the 60-GHz communication systems. Different from interference analysis in conventional low-frequency networks, interference in mmW bands is mainly caused by concurrent directional communications links. It is found that multi-hop MAC protocol based on these models is effective to maintain high network utilization with low overhead \cite{jsac09}.

Also, other prior work studied coexistence of Wi-Fi and cellular in lower-frequency bands. The coexistence of LAA-LTE and Wi-Fi in indoor environments was studied \cite{icc15}. This experimental study performed implementation of LAA-LTE and Wi-Fi. The findings are the facts that (i) a small bandwidth of LAA-LTE (1.4/3/5/10MHz) causes a greater impact on the Wi-Fi performance, and (ii) LAA-LTE signals with LAA-LTE can trigger channel busy indication of CS/CCA in Wi-Fi. However, the paper did not provide thorough technical rationale behind the findings. Another relevant study focused on the coexistence between Wi-Fi and small-cell LTE \cite{commga15}. A novel network architecture for LTE/LTE-A small cells was proposed to exploit the unlicensed spectrum already used by Wi-Fi systems. The interference avoidance scheme was presented to mitigate the interference between Wi-Fi and LTE/LTE-A systems when both operating in the same unlicensed spectrum.

More technical approaches on inter-technology coexistence were presented \cite{lett}-\cite{hindawi}\cite{wcnc}. A spatio-temporal analysis on the military radar-Wi-Fi coexistence in 3.5 GHz band was studied \cite{lett}. Exploiting the fact that a military radar ``rotates'' in a fixed revolution rate, the Wi-Fi was proposed to transmit while a radar beam faces to the other directions. The study measured (i) how much performance a Wi-Fi achieves and (ii) how much interference leaks into the radar. In the same 3.5 GHz band, another study assumed Long-Term Evolution (LTE) as the secondary system \cite{hindawi}. The cellular technology based on orthogonal frequency-division multiplexing (OFDM) adopted a larger inter-subcarrier spacing to overcome the pulsed interference from a coexisting radar. Based on the assumption that the 3.5-GHz coexistence requires a spectrum access system (SAS), another relevant study analyzed the impact of ``imperfect sensing'' performed at a SAS on the performance of coexistence \cite{wcnc}.

Distinguished from the prior work, this paper presents the following contributions:
\begin{enumerate}
\item It addresses inter-technology coexistence in the 60 GHz band. The unique communications characteristics necessitates thorough study on the coexisting wireless systems--\textit{i.e.}, Wi-Fi and 5G. This paper focuses on modeling the data rates achieved by Wi-Fi and 5G in a number of possible scenarios.
\item It focuses on the applicability of communications in the 60 GHz to HA/DR operations.
\end{enumerate}

\vspace{0.2 in}
\section{Specific Technical Challenges}\label{sec_challenges}
The Wireless Broadband Alliance (WBA) is one of the organizations that are the most actively leading the discussions on the Wi-Fi-cellular coexistence \cite{wba}. The organization identified gaps regarding the coexistence of technologies, convergence of services, and certification and operator guidelines. As 5G and Wi-Fi continue to shape and be shaped by each other, WBA stakeholders will need to fast-track solutions for items including seamless authentication; aggregated access; multi-access edge computing (MEC)-enabled service delivery; extreme real-time communications; network slicing; high-speed transport use cases; roaming for non-3GPP subscription identifiers, and keying hierarchies--all of which are explored in more detail within the white paper.

Solving for these issues will enable a range of new use cases, going beyond the consumer applications of the past into new verticals like vehicle-to-vehicle (V2V) and vehicle-to-everything (V2X), smart factories, high-density cities, and public safety solutions. In a world where everything is smart and connected, 5G will enable consumers, businesses, municipalities, and industries to unlock the power of internet of things (IoT).

Much work remains to be done. This paper identifies key technical challenges in achieving coexistence and interoperability of Wi-Fi and 5G in the 60 GHz band. First, deployment-related challenges can be listed as
\begin{itemize}
\item There is a need for recommendations to ensure that Wi-Fi selection does not produce a bad user experience \textit{e.g.} hanging on to slow/distant service set identifiers (SSIDs) for too long or moving to an AP that is congested or has poor Quality of Service metrics. This is particularly the case if auto-network selection is used in implementations which makes it hard for the end-user to override bad behavior.
\item Efficient prioritization of SSID selection in a multi-service environment is important for monetization strategies; however, current deployments are not achieving this.
\item In current set of devices, the entity responsible for ``best connected user experience'' is split between user equipment (UE) manufacturer, service provider and end-user. The policy elements need to reflect this split and work in conjunction with all these components.
\item There are no access network discovery and selection function (ANDSF) deployments as the network elements supporting current ANDSF standards are not available, nor is there any support for ANDSF policies in any of the commercial or popular operating systems (OSs) or client platforms.
\end{itemize}

The identification of challenges goes on to application-related considerations:
\begin{itemize}
\item There is no interoperability among ANDSF / Hotspot (HS) 2.0 vendors; roaming / visiting policy is not well defined in ANDSF; and deployment by operators owning both cellular and Wi-Fi networks is rare.
\item The industry knowledge base does not have a common understanding of the aims, methods and ways of implementation of different policy components.
\item Provide the best user experience given the available air links and conditions. However, the user needs to have the final say in influencing selection, as their ``best'' user experience may involve an inferior connection over a less costly access network (User preferred connections).
\item Order of connection preference should be HS 2.0, then 802.1X, then Open (assuming all are `public' and not `remembered').
\item Maintain stable connections (\textit{e.g.} multipath transmission control protocol (MP-TCP), LTE, Wi-Fi and seamless hand-over).
\item Prevent Ping-Pong hand-over (\textit{e.g.} LTE - Wi-Fi handover).
\end{itemize}

Summarizing and taking a systematic point of view, the key technical challenges in achieving the coexistence between Wi-Fi and 5G can be summarized as follows:
\begin{itemize}
\item Both coexisting systems have complicated traffic patterns;
\item Both coexisting systems are mobile;
\item Both coexisting systems require very high data rates--\textit{i.e.}, higher than several Gbps
\end{itemize}

\vspace{0.2 in}
\section{Coexistence Analysis}\label{sec_coexistence}
For analysis, this paper assumes three representative coexistence scenarios between Wi-Fi and 5G in the 60 GHz band as illustrated in Fig. \ref{fig_pairs}:
\begin{itemize}
\item \textbf{Scenario 1:} Wi-Fi only
\item \textbf{Scenario 2:} 5G only
\item \textbf{Scenario 3:} 5G and Wi-Fi coexisting
\end{itemize}

\subsection{Small-Scale Environment}
Consider a network that is built by two pairs of transmitters (Tx's) and receivers (Rx's). In each system of Wi-Fi and 5G, every Tx is assumed to serve only one Rx. Fig. \ref{fig_pairs} illustrates three coexistence scenarios that represent operations of 5G and Wi-Fi technologies in the 60 GHz band.

In Scenario 1, the data rate achieved from a Wi-Fi AP to its Rx can be written as
\begin{align}
R_W = \alpha B_{W} \log (1+SNR_W)
\end{align}
where $\alpha$ indicates the portion of an active transaction time. Also, $B_{W}$ gives the bandwidth that a Wi-Fi system utilizes, and $SNR_W$ is the signal to noise on the recipient in this sub-case. Note that $\log$ is with base of 2.

Scenario 2 is composed of the two pairs of transceivers are 5G. The data rate can be calculated as
\begin{align}
R_{5G} &= B_{L} \log(1+SINR_{\text{l-to-u}})\nonumber\\
&{\rm{~~~~~~~~~~~~}}+ \alpha\rho B_{U} \log(1+SNR_{5G,U})
\end{align}
where $SINR_{\text{l-to-u}}$ is the 5G NR-U receiver's signal to interference plus noise ratio; the interference comes from the `licensed' segment of the 5G system. In contrast, $SNR_{\text{u}}$ regards the noise generated in the `unlicensed' link of a 5G system. $B_{5G}$ indicates the bandwidth of a 5G link.

Scenario 3 represents the inter-system coexistence between Wi-Fi and 5G. The data rate achieved by Wi-Fi can be written as
\begin{align}
R'_{W} &= \alpha B_{W} \log (1+SNR_{W})\nonumber\\
&{\rm{~~~~~~~~~~~~}}+ (1-\alpha) B_{W} \log(1+SINR_{\text{5Gu-to-W}})
\end{align}
where $SINR_{\text{5Gu-to-W}}$ consists of interference generated by the 5G unlicensed Tx. Also, the data rate of 5G can be obtained as
\begin{align}
R'_{5G} &= B_{L} \log (1+SINR_{l-to-u})\nonumber\\
&{\rm{~~~~~~~~~~~~}}+ (1-\alpha) \rho B_{U} \log(1+SINR_{\text{W-to-5Gu}}).
\end{align}

\subsection{Large-Scale Environment}
Now, we extend our analysis to a scenario with a larger number of nodes. For each system of Wi-Fi and 5G, the nodes are dropped following Poisson point process (PPP) following the intensities of $\lambda_{W}$ and $\lambda_{5G}$, respectively. For a PPP, the probability that $N$ nodes exist in an area $\mathcal{A}$ given an intensity $\lambda$ is written as \cite{haenggi05}
\begin{align}\label{eq_ppp}
P(N|\lambda)=\frac{\left( \lambda|\mathcal{A}| \right)^N}{N!}e^{-\lambda|\mathcal{A}|},\:\: N=0,1,2,...,\infty
\end{align}
where $|\mathcal{A}|$ denotes the area of $\mathcal{A}$.

Based on Eq. (\ref{eq_ppp}), the data rates achieved by Wi-Fi and 5G without coexistence in Scenarios 1 and 2, respectively, are given by
\begin{align}
R_{W} &= \frac{1}{f(N)}B_{W} \log(1+SNR_{W})\\
R_{5G} &= \frac{N}{f(N)}B_{5G} \log(1+SINR_{\text{l-to-u}})\nonumber\\
&{\rm{~~~~~~~~~~~~}}+ \frac{1}{f(N)}B_{5G}\log(1+SNR_{5G})
\end{align}
where $f(N)$ is a component of transmitters density which shows the number of transmitters $N$.

The data rates achieved in Wi-Fi and 5G with interference from each other are given by
\begin{align}
R'_{W} &= \frac{2}{f(N)}B_{W}\log(1+SNR_{W})\nonumber\\
&{\rm{~~~~~~~~~~~~}}+ \frac{2}{f(N)}B_{W}\log(1+SINR_{\text{5Gu-to-W}})\\
R'_{5G} &= \frac{N}{2f(N)}B_{L}\log(1+SINR_{\text{l-to-u}})\nonumber\\
&{\rm{~~~~~~~~~~~~}}+ \frac{1}{f(N)}B_{U}\log(1+SNR_{5G}).
\end{align}

\begin{table}[t]
\centering
\caption{Simulation parameters}
\label{table_parameters}
\begin{tabular}{|c|c|}
\hline
Parameters & Values \\ \hline
Time fraction ($\alpha$) & [0,1] \\ \hline
Path loss model & Free-space path loss \cite{fspl} \\ \hline
Bandwidth & 2.16 GHz (single channel) \\ \hline
Tx/Rx intensity ($\lambda_{W}$, $\lambda_{5G}$) & $10^{-4}$ $m^{-2}$ \\ \hline
Area of experiment & 100 $m^2$ \\ \hline
Tx power on 5G AP & 27 dBm \\ \hline
Tx power on Wi-Fi AP & 27 dBm \\ \hline
Noise figure, NF & 1.5 dB \\ \hline
\end{tabular}
\end{table}

\begin{figure*}[t]
\centering
\includegraphics[width = 0.6\linewidth]{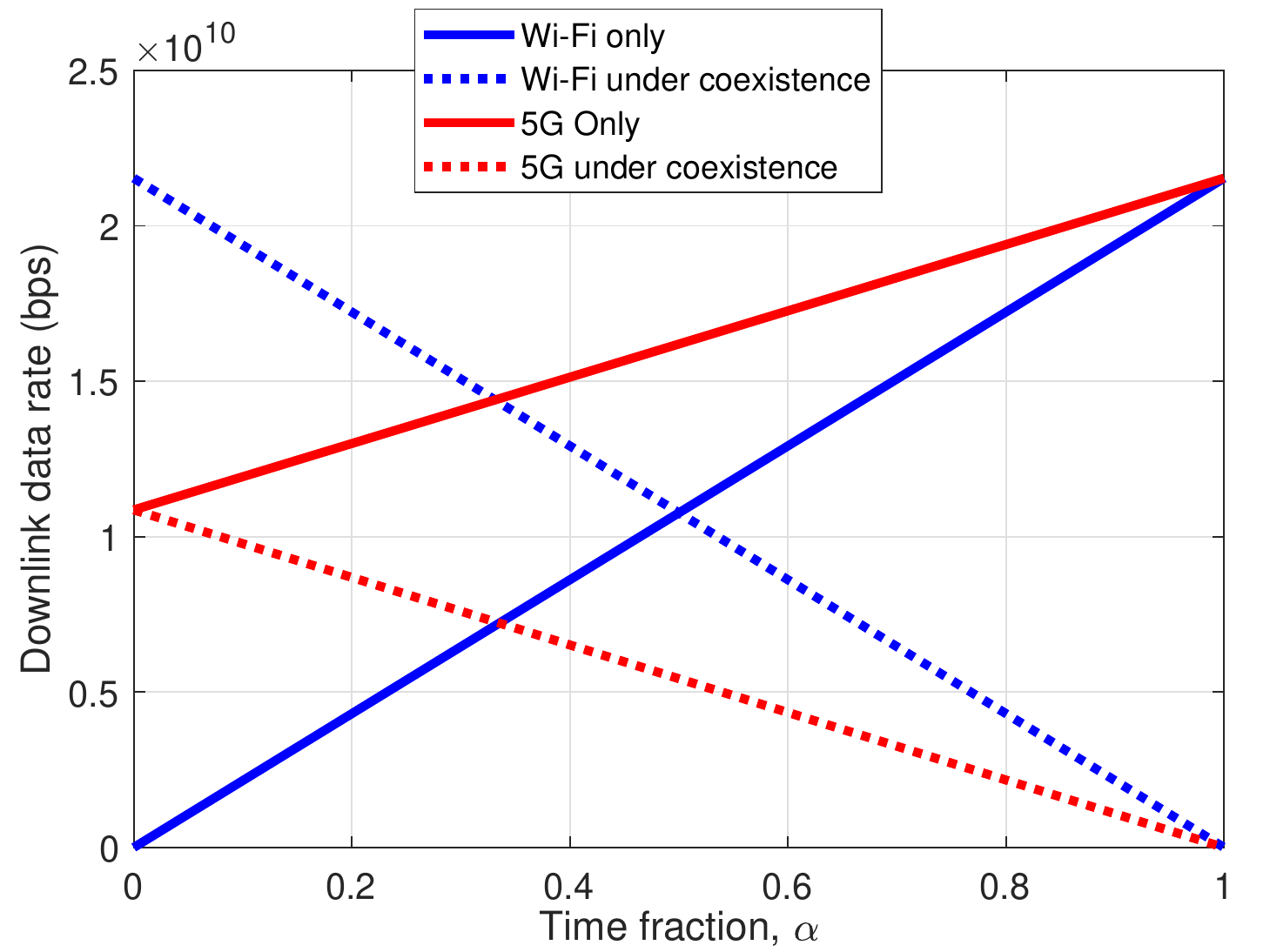}
\caption{Downlink data rate according to the time fraction, $\alpha$}
\label{fig_alpha}
\vspace{0.4 in}
\centering
\includegraphics[width = 0.6\linewidth]{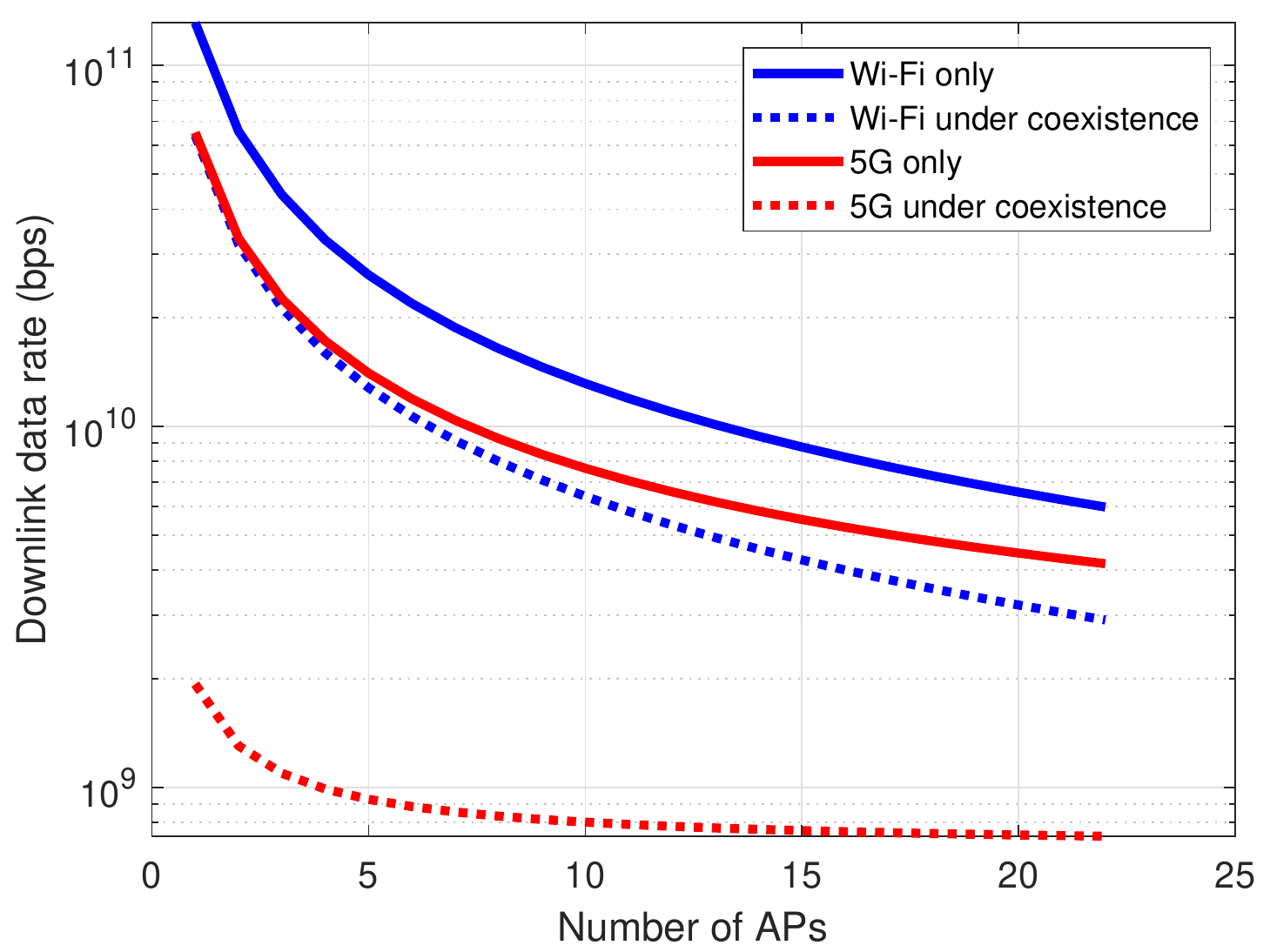}
\caption{Download data rate according to the node density}
\label{fig_density}
\end{figure*}

\vspace{0.2 in}
\section{Results and Discussion}
MATLAB simulations are performed in order for evaluation of the data rates formulated in Section \ref{sec_coexistence}. The key parameters are summarized in Table \ref{table_parameters}.

\subsection{Data Rate according to Time Fraction}
Fig. \ref{fig_alpha} shows downlink data rates achieved in the network settings shown in Fig. \ref{fig_pairs}. Recall that $\alpha$ denotes a fraction of the time taken for a Tx-Rx pair of a wireless system--\textit{viz.}, either Wi-Fi or 5G. The key observations are as follows:
\begin{itemize}
\item For both Wi-Fi and 5G, without coexistence considered, a larger time fraction leads to a higher data rate. The increase is slower in 5G due to the internal interference from the `licensed' segment of the system.
\item In contrast, when coexistence is considered, the data rate is decreased with a larger value of time fraction. The rationale is as follows. A larger value of time fraction implies a longer occupancy of a certain Tx-Rx pair, which in turn acts as a greater interference to the other Tx-Rx pair. Therefore, the system-level downlink data rate is decreased due to such a higher interference caused by a larger $\alpha$.
\end{itemize}

\subsection{Data Rate according to Node Density}
Fig. \ref{fig_density} demonstrates downlink data rates obtained according to the number of APs. The following observations are found:
\begin{itemize}
\item Wi-Fi achieves higher data rates than 5G since 5G undergoes the internal interference between unlicensed and licensed segments.
\item Coexistence with the other technology (between Wi-Fi and 5G) degrades the data rates.
\item A larger number of APs lowers the data rates by increasing the level of bandwidth contention.
\end{itemize}

Notice that the key technical challenge in coordination of coexistence between 5G and Wi-Fi is the dissimilarity in being ``synchronous.'' Specifically, the performance degradation due to interference can be far more severe in Wi-Fi than 5G, because a Wi-Fi system is asynchronous. Assuming a distributed coordination function (DCF), a Wi-Fi station (STA) needs to hold transmission of a packet until the channel becomes idle. Even after having the channel idle, the STA needs to hold for its assigned backoff time before transmission. The problem is that during such a relatively long hold time, a 5G Tx could start another transmission, which will make the channel busy again. As such, as an asynchronous system, the Wi-Fi keeps a handicap over the 5G. We suggest this as a direction of modifying the Wi-Fi to better fit to coexistence with the 5G.

\subsection{Insights on HA/DR Applications}
As mentioned earlier in Section \ref{sec_intro}, the data rate acts as the key performance indicator in delivering mission-critical information and data during a HA/DR activity. For example, a rescue of victims isolated in a disaster area would require a certain level of data rate to support real-time duplex video communications. For emergency communications applied in such HA/DR scenarios, one expects \textit{high reliability}, \textit{high availability}, and \textit{low latency}. For the three performance requirements, it is known that a minimum data rate of 10 Mbps for both uplink and downlink is needed \cite{lett}\cite{hindawi}.

Figs. \ref{fig_alpha} and \ref{fig_density} suggest that despite the coexistence with each other, both Wi-Fi and 5G technologies are capable of achieving higher data rates than the aforementioned required data rate. It implies that both technologies are feasible for HA/DR missions.

\section{Conclusion and Future Work}
This paper investigated the feasibility of Wi-Fi and 5G technologies at 60 GHz for HA/DR applications. Establishment of an emergency communications system in the 60 GHz spectrum can bring a huge benefit in stable management of HA/DR activities for two reasons: (i) no license nor paid subscription is needed, which enables higher availability; (ii) a very large bandwidth is available, which significantly increases the reliability and reduces latency. Our simulation results show that both of Wi-Fi and 5G can achieve data rates ranging from a few to tens of Gbps, which exceed the requirements for most of the HA/DR missions. As such, it proves that both wireless technologies at 60 GHz have a potential to fulfill high-data-rate HA/DR missions such as real-time video streaming.

As future work, we seek to make the system model more realistic. It will figure out an exact amount of degradation in the data rate, which will consequently suggest practical HA/DR deployment scenarios.



\begin{thebibliography}{99}
\setlength{\parskip}{0.5 em}
\bibitem{fema} FEMA, ``Disaster emergency communications,'' [Online]. Available: \url{https://www.fema.gov/disaster-emergency-communications}, Apr. 2018, Accessed on: May 13, 2019.

\bibitem{fcc} FCC, ``Report and order and further notice of proposed rulemaking,'' FCC-16-89A1, Jul. 2016.

\bibitem{lu2017downlink} X. Lu, M. Lema, T. Mahmoodi, and M. Dohler, ``Downlink data rate analysis of {5G-U} ({5G} on unlicensed band): coexistence for {3GPP 5G} and {IEEE} 802.11ad {WiGig},'' in \textit{Proc. European Wireless Conf.}, 2017.

\bibitem{yang06} L. Lily Yang, ``60GHz: opportunity for gigabit WPAN and WLAN convergence,'' in \textit{Proc. ACM SIGCOMM 2006}.

\bibitem{rappaport} C. Park and T. S. Rappaport, ``Short-range wireless communications for next-generation networks: UWB, 60 GHz millimeter-wave WPAN, and ZigBee,'' \textit{IEEE Wireless Commun.}, Aug. 2007.

\bibitem{jsac09} S. Singh, F. Ziliotto, U. Madhow, E. Belding, and M. Rodwell, ``Blockage and directivity in 60 GHz wireless personal area networks: from cross-layer model to multihop MAC design,'' \textit{IEEE J. Sel. Areas Commun.}, vol. 27, no. 8, 2009.

\bibitem{icc15} Y. Jian, C.-F. Shih, B. Krishnaswamy, and R. Sivakumar, ``Coexistence of Wi-Fi and LAA-LTE: experimental evaluation, analysis and insights,'' in \textit{Proc. IEEE ICC 2015}.

\bibitem{commga15} H. Zhang, X. Chu, W. Guo, and S. Wang, ``Coexistence of Wi-Fi and heterogeneous small cell networks sharing unlicensed spectrum,'' \textit{IEEE Commun. Mag.}, Mar. 2015.

\bibitem{lett} S. Kim and C. Dietrich, ``Coexistence of outdoor Wi-Fi and radar at 3.5 GHz,'' \textit{IEEE Wireless Commun. Lett.}, vol. 6, iss. 4, Aug. 2017.

\bibitem{hindawi} S. Kim, J. Choi, and C. Dietrich, ``PSUN: an OFDM - pulsed radar coexistence technique with application to 3.5 GHz LTE,'' \textit{Hindawi Mobile Inform. Syst. J.}, Jul., 2016.

\bibitem{wcnc} S. Kim, J. Choi, and C. Dietrich, "Coexistence between OFDM and pulsed radars in the 3.5 GHz band with imperfect sensing," in Proc. IEEE Wireless Communications and Networking Conference (WCNC) 2016.

\bibitem{wba} Wireless Broadband Alliance (WBA), ``Convergence of cellular and next gen Wi-Fi networks – ANDSF and HS 2.0 policies,'' \textit{Policy Networking}, Apr. 2017.

\bibitem{haenggi05} M. Haenggi, ``On distances in uniformly random networks,'' \textit{IEEE Trans. Inf. Theory}, vol. 51, no. 10, Oct. 2005.

\bibitem{fspl} S. Kim, E. Visotsky, P. Moorut, K. Bechta, A. Ghosh, and C. Dietrich, ``Coexistence of 5G with the incumbents in the 28 and 70 GHz bands,'' \textit{IEEE J. Sel. Areas Commun.}, vol. 35, iss. 8, Aug. 2017.
\end{thebibliography}
\end{document}